\documentclass[11pt]{article}
\usepackage{amsmath,amsfonts,amssymb,amsthm}
\usepackage{graphicx}
\usepackage{tikz}
\usepackage{enumerate}
\usepackage[hyphens]{url}
\usepackage{xcolor}
\usepackage{url}
\usepackage{caption}
\usepackage{subcaption}
\usepackage{tcolorbox}
\usepackage{hyperref}
\usepackage{chemarr}
\usepackage[margin=3.5cm]{geometry}
\usepackage{capt-of}
\usepackage{authblk}
\usepackage{booktabs}
\usepackage{soul}
\theoremstyle{plain}
\newtheorem{theorem}{Theorem}

\theoremstyle{definition}
\newtheorem{definition}{Definition}[section]
\newtheorem{example}[theorem]{Example}
\newtheorem{remark}{Remark}[section]

\newcommand\R{\mathbb{R}}

\definecolor{darkgreen}{rgb}{0.0,0.7,0.0}

\usepackage{tikz}
\usetikzlibrary{patterns}
\usetikzlibrary{matrix, arrows, decorations.pathmorphing}
\usepackage{tikz-cd}
\tikzset{commutative diagrams/.cd}
\usepackage{framed,lipsum}
\usepackage{float}
\usepackage{pgfplots}
\usetikzlibrary{patterns}
\usetikzlibrary{intersections}
\usetikzlibrary{positioning, shadows, arrows}
\tikzstyle{every node}=[anchor=west, minimum height=3em]
\usetikzlibrary{decorations.pathmorphing}

\newcommand{\mS}{\mathcal{S}}
\newcommand{\mC}{\mathcal{C}}
\newcommand{\mR}{\mathcal{R}}
\newcommand{\SCR}{(\mathcal{S}, \mathcal{C}, \mathcal{R})}
\newcommand{\SCRk}{(\mathcal{S}, \mathcal{C}, \mathcal{R}, \boldsymbol{k})}
\newcommand{\bk}{\boldsymbol{k}}
\newcommand{\bx}{\boldsymbol{x}}

\usepackage{enumitem}

\renewcommand{\arraystretch}{1.3}  

\renewcommand{\arraystretch}{1.3}  

\begin{document}

\title{Implementation of Support Vector Machines using Chemical Reaction Networks}

\author[1]{Amey Choudhary}
\author[2]{Abhishek Deshpande}
\author[3]{Jiaxin Jin}
\affil[1,2]{\small Center for Computational Natural Sciences and Bioinformatics, International Institute of Information Technology, Hyderabad} 
\affil[3]{\small Department of Mathematics, University of Louisiana at Lafayette}

\maketitle

\begin{abstract}

Can machine learning algorithms be implemented using chemistry? We demonstrate that this is possible in the case of support vector machines (SVMs).
SVMs are powerful tools for data classification, leveraging Vapnik–Chervonenkis theory to handle high-dimensional data and small datasets effectively. In this work, we propose a chemical reaction network scheme for implementing SVMs, utilizing the steady-state behavior of reaction network dynamics to model key computational aspects of SVMs. 
This approach introduces a novel biochemical framework for implementing machine learning algorithms in non-traditional computational environments.
\end{abstract}

\section{Introduction}

Living cells respond to external stimuli through mechanisms, such as gene regulatory networks. Chemical reaction network theory hopes to construct complex circuits, such as oscillators and switches, from simple building blocks. Recently, these designs have enabled the development of complicated biological circuits for real-world applications. Our work aims to enable computation in living environments, where modern computers are currently unable to operate.

Chemical reaction networks provide a robust framework for biomolecular interactions and can be realized through DNA strand displacement reactions~\cite{simmel2019principles}, providing a physical basis in cellular environments. This makes them a powerful ``programming language'' for molecular computation, benefiting from DNA’s sequence specificity and high-density information storage. However, programming reaction networks for more sophisticated tasks, such as training and testing, remains a significant challenge.

The idea of using reaction networks for performing computation dates back to 1994, when Adleman solved a seven-node Hamiltonian path problem using DNA strands~\cite{adleman1994molecular}. Winfree and Qian later developed DNA strand displacement schemes capable of realizing increasingly complex molecular circuits~\cite{cherry2018scaling,qian2011neural}.
More recently, reaction networks have been used to implement various machine learning algorithms. 
In particular, Stojanovic et al.\cite{pei2010training} constructed reaction networks for automating decision tree implementations, while Gopalkrishnan et al.\cite{gopalkrishnan2016scheme,virinchi2017stochastic,virinchi2018reaction} designed schemes for maximum likelihood estimation and Boltzmann machine simulation~\cite{poole2017chemical,poole2025autonomous,poole2022detailed}. Reaction network implementations of neural network dynamics have also been demonstrated~\cite{anderson2021reaction,fan2023automatic,kang2024noise}.

In this paper, our focus is on \emph{Support Vector Machines (SVMs)}. Introduced by Vapnik and Chervonenkis in the 1960s, SVMs are supervised classification algorithms. The essential idea of SVM is to construct a hyperplane that maximally separates data points from different classes, creating the largest possible margin. While SVMs are highly effective for linearly separable data, nonlinear classification requires the use of \emph{kernel functions} to map data onto higher-dimensional spaces. This is also called the \emph{kernel trick} for SVM~\cite{guyon1992automatic}.

We design a reaction network scheme to implement a soft-Margin SVM, leveraging biochemical processes for machine learning tasks. Our reaction network-based model is capable of (1) loading input data points in batches, (2) performing inference, (3) backpropagating the classification loss, and (4) updating weights via gradient descent.
To ensure these operations are executed sequentially, we employ oscillating molecular species to achieve time-scale separation. We validate our design by simulating the reaction network on synthetic datasets using Python and demonstrate effective convergence, with weight updates closely matching those obtained from standard SVM implementations. This work introduces a novel biochemical framework for executing machine learning algorithms in non-traditional computational environments, broadening the scope of reaction network applications in synthetic biology.

\textbf{Structure of the paper: }
The paper is organized as follows. 
In Section~\ref{sec:reaction_networks}, we introduce reaction networks and recall their properties. Section~\ref{sec:support_vector_machines} introduces support vector machines and describes their functioning as an optimization problem. 
In Section~\ref{sec:modules}, we introduce the concept of dual-rail encoding to handle negative weights and biases. Further, we describe basic modules like addition, subtraction, multiplication, comparison, and approximate majority that will be used as building blocks in the reaction network that simulates the SVM. 
In section ~\ref{sec:hopf_clock}, we describe our implementation of our molecular clock using \textit{Hopf} reactions.
In Section~\ref{sec:assignment}, we describe the assignment module that is used for loading training in batches.
Sections~\ref{sec:feedforward} and \ref{sec:learning} describe the reaction network that simulates the feedforward and 
backpropagation components of the SVM, respectively.
In Section~\ref{sec:analysis}, we present and validate the results for our reaction network and compare the values  of weights and biases obtained using reaction networks and those obtained using the traditional SVM implementations. To validate these results, we also compare their trajectories as a function of time (epochs in our case) and find that they are in close agreement. 
Section~\ref{sec:discussion} summarizes our contributions and outlines directions for future work.

\section{Chemical Reaction Networks}
\label{sec:reaction_networks}

\begin{definition}[\cite{feinberg}]

A Chemical Reaction Network (reaction network) is a triple $\SCR$ consisting of a finite set of \textit{species} $\mS = \{X_1, \dots, X_n \}$, a set of \textit{complexes} $\mC = \{ C_1, \dots, C_s \}$, and a set of \textit{reactions} $\mR = \{ R_1, \dots, R_r \}$. 
Each reaction describes an interaction between species given by:
\begin{equation} \notag
R_j : \sum_{i=1}^n \alpha_{ij} X_i \rightarrow \sum_{i=1}^n \beta_{ij} X_i
\ \text{ with } j = 1, \dots, r,
\end{equation}
where the left-hand side (\textit{reactant complex}) and the right-hand side (\textit{product complex}) are linear combinations of species.
For each reaction $R_j$, the non-negative integer coefficients $ \alpha_{ij} $ and $ \beta_{ij} $ represent the \textit{stoichiometric coefficients} of species $ X_i $ in the reactant and product complexes, respectively.
\end{definition}

\begin{example} 
\label{ex:crnt} 

Consider the reaction network $\SCR$ shown in Figure \ref{fig:ex1}. It consists of three species:  
\[
\mS = \{X_1, X_2, X_3 \},
\]  three complexes:  
\[
\mC = \{ X_1 + X_2, \ 2X_2, \ X_3 \},
\]  
and four reactions:  
\[
\mR = \{ X_1 + X_2 \to X_3, \quad  
2X_2 \to X_1 + X_2, \quad  
2X_2 \to X_3, \quad  
X_3 \to X_1 + X_2 \}.
\]
\begin{figure}[!ht] 
\begin{center}
\begin{tikzpicture}
		\node (1) at (-3,0) {$\bullet$};
		\node (3) at (0,2) {$\bullet$};
		\node (2) at (3,0) {$\bullet$};
		\node [left=1pt of 1] {$X_1 + X_2$};
		\node [right=1pt of 2] {$2X_2$};
		\node (x3) at (0,2.5) {$X_3$};
		\draw [{->}, -{stealth}, thick, transform canvas={yshift=2pt}] (1) -- (3) node [midway, above] {};
		\draw [{->}, -{stealth}, thick, transform canvas={yshift=-0pt}] (2) -- (1) node [midway, below] {};
		\draw [{->}, -{stealth}, thick, transform canvas={xshift=1.5pt}] (2) -- (3) node [midway, right=6pt] {};
		\draw [{->}, -{stealth}, thick, transform canvas={yshift=-2pt}] (3) -- (1) node [midway, left] {};
\end{tikzpicture}
\end{center}
\caption{A reaction network $\SCR$ with three species, three complexes, and four reactions.}
\label{fig:ex1}
\end{figure}
\end{example}

One of the most widely used models in reaction network studies is based on \textit{mass-action kinetics} \cite{guldberg1864studies}. In this framework, the reaction rate is determined by the product of the reactant concentrations, each raised to its corresponding stoichiometric coefficient. The detailed definition is provided below.

\begin{definition}[\cite{feinberg}]

Given a reaction network $\SCR$, each reaction $R_j$ is assigned a positive constant $k_j$ (\textit{reaction rate constant}) for $1 \leq j \leq r$. Let $\bk = (k_1, \dots, k_r) \in \mathbb{R}_{>0}^{r}$ denote the \textit{reaction rate vector}. The \textit{mass-action system} generated by $\SCRk$ is given by:
\begin{equation} \notag
\frac{d x_i (t)}{dt} = \sum_{j=1}^r k_j \prod_{l=1}^n x_l^{\alpha_{lj}} (\beta_{ij} - \alpha_{ij})
\ \text{ with } i = 1, \dots, n,
\end{equation}
where $\bx (t) = (x_1(t), \dots, x_n(t))$ represents the concentrations of species $X_1, \dots, X_n$ at time $t$.
\end{definition}

For example, recall the reaction network from Figure \ref{fig:ex1} (see Example \ref{ex:crnt}). Given the reaction rate vector $\bk = (k_1, \dots, k_4)$, the reaction rates are assigned as follows:
\[
X_1 + X_2 \xrightarrow{k_1} X_3 \qquad  
2X_2 \xrightarrow{k_2} X_1 + X_2 \qquad  
2X_2 \xrightarrow{k_3} X_3 \qquad  
X_3 \xrightarrow{k_4} X_1 + X_2
\]
Under mass-action kinetics, the associated dynamical system is
\begin{equation} \notag
\begin{split}
\frac{\mathrm{d}\bx}{\mathrm{d} t}
& = k_{1} x_1 x_2 \begin{pmatrix} -1 \\ -1 \\ 1 \end{pmatrix}
+ k_{2} x_2^2 \begin{pmatrix} 1 \\ -1 \\ 0 \end{pmatrix} 
+ k_{3} x_2^2 \begin{pmatrix} 0 \\-2\\1 \end{pmatrix}
+ k_{4} x_3 \begin{pmatrix} 1 \\ 1 \\ -1 \end{pmatrix}
\\& = \begin{pmatrix} 
- k_{1} x_1 x_2 + k_{2} x_2^2 + k_{4} x_3 
\\ -k_{1} x_1 x_2 + (k_{2} - 2 k_{3})x_2^2 + k_{4} x_3
\\ k_{1} x_1 x_2 + k_{3} x_2^2 - k_{4} x_3
\end{pmatrix}.
\end{split}
\end{equation}

Reaction networks offer a robust framework for implementing molecular-level computations since the mass-action systems generated by them have polynomial right-hand sides. 

In the following, we introduce two key concepts in reaction networks: catalytic and non-catalytic species, which will be frequently used in Section~\ref{sec:modules}.

\begin{definition}[\cite{feinberg}]

Let $\SCR$ be a reaction network with reactions $R_1, \dots, R_r \in \mR$, where each reaction has the form
\[
R_j: \sum_{i=1}^n \alpha_{ij} X_i \rightarrow \sum_{i=1}^n \beta_{ij} X_i
\]
\begin{enumerate}
\item[(a)] A species $X_i \in \mS$ is said to be \emph{catalytic} if the number of species $ X_i$ in the reactant and product complexes remains the same for every reaction. Specifically, $X_i$ is a catalytic species if and only if $\alpha_{ij} = \beta_{ij}$ for every $1 \leq j \leq r$.

\item[(b)] A species $X_i \in \mS$ is said to be a \emph{non-catalytic} species if it is not catalytic, that is, there exists a reaction $R_j$ such that $\alpha_{ij} \neq \beta_{ij}$. 
\end{enumerate}
\end{definition}

\begin{example}

Consider the following reaction network:
\[
\emptyset \to X_1 + X_2 \qquad
X_1 + X_3 \to X_3
\]
In this example, the reaction network consists of three species, $\mS = \{ X_1, X_2, X_3 \}$. Among them, $X_1$ and $X_2$ are non-catalytic, while $X_3$ is a catalytic species.
\end{example}

\section{Support Vector Machines}\label{sec:support_vector_machines}

Support Vector Machines (SVMs) \cite{cortes1995support,vapnik1997support} are supervised machine learning algorithms. SVMs have attracted widespread attention in machine learning, data mining, and pattern recognition tasks. In particular, they are particularly useful in binary classification problems. 

SVM works on the principle of generating a surface that separates data into classes by maximizing the distance between itself and the data points. In particular, SVM assigns the data points labels by finding the \emph{optimal hyperplane}. This hyperplane corresponds to one that is at the maximum distance from the closest point of any class. Given a new data point, it can be classified and given labels according to which side of the hyperplane it will be in feature space.

If the data cannot be separated by such a hyperplane, then the SVM finds the hyperplane as in the hard margin case, with the caveat that a penalty is added each time a point crosses the margin. These SVMs are referred to as soft margin SVMs.

\begin{figure}
    \centering
    \includegraphics[width=0.75\linewidth]{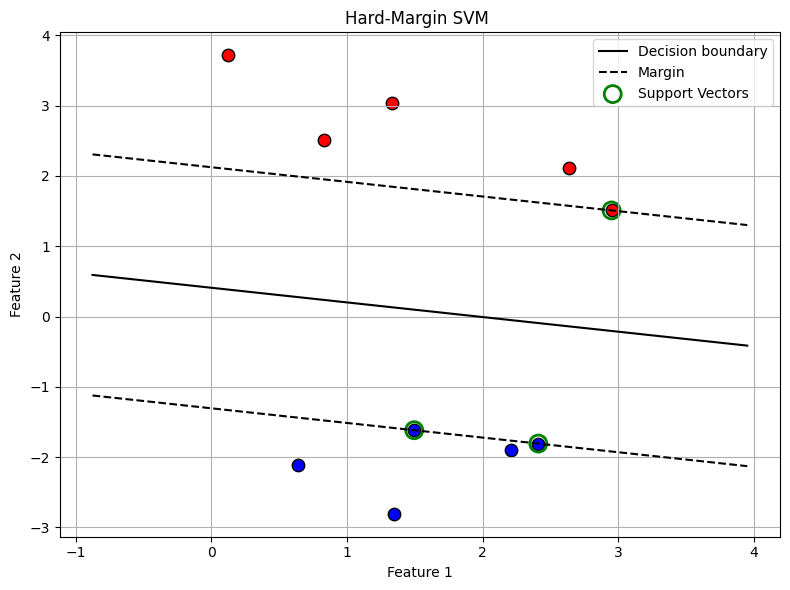}
    \caption{Visualization of Hard Margin SVM for two-dimensional input and class label $\in \{1, -1\}$; red points denote class $1$ and blue points denote class $-1$.}
    \label{fig:enter-label}
\end{figure}

\subsection{Hard-Margin SVM Formulation}

Given a set of data points (which are feature vectors) that belong to either of the two classes, one needs to decide to which of the two classes a new data point belongs. More formally, given a dataset \( \{(\bx^{(i)}, y^{(i)})\}_{i=1}^N \), where \( \bx^{(i)} \in \mathbb{R}^d \) are the feature vectors and \( y^{(i)} \in \{1, -1\} \) are the binary class labels, the goal is to find a weight vector \( \mathbf{w} \in \mathbb{R}^d \) 
such that it maximizes the distance of its nearest point for each class and bias \( b \in \mathbb{R} \) such that:
\begin{equation}\label{eq:max_margin}
y^{(i)} (\mathbf{w} \cdot \bx^{(i)} + b) \geq 1 
\quad \text{for all } 1 \leq i \leq N.
\end{equation}
The optimization problem for a \textit{hard-margin SVM} is defined as:
\begin{equation}\label{eq:hard_svm_opt}
    \min_{\mathbf{w}, b} \frac{1}{2} \|\mathbf{w}\|^2,
\end{equation}
subject to the constraint that all training points satisfy Equation \eqref{eq:max_margin}.
This objective aims to maximize the margin, which is inversely proportional to $\|\mathbf{w}\|$.

Figure~\ref{fig:enter-label} shows a hard-margin SVM for a two-dimensional input and class label taking values $-1$ and $1$. 

\subsection{Theoretical Justification}

While the hard-margin SVM provides an ideal formulation for linearly separable data, directly solving this constrained optimization problem can be complex. In practice, we adopt an unconstrained formulation using \textit{hinge loss} and \textit{L2 regularization}, which corresponds to a \textit{soft-margin SVM}:
\begin{equation} \notag
\min_{\mathbf{w}, b} \quad \frac{\lambda}{2} \|\mathbf{w}\|^2 + \sum_{i=1}^N \max(0, 1 - y^{(i)}(\mathbf{w} \cdot \bx^{(i)} + b)),
\end{equation}
where $\lambda > 0$ is the regularization parameter. This soft-margin approach is more robust as it allows for margin violations, penalizing them through the hinge loss. A key insight is that when the data is linearly separable, the hinge loss term becomes zero for all training points, effectively reducing the soft-margin objective back to the original hard-margin problem in Equation \eqref{eq:hard_svm_opt}.

The iterative training algorithm we employ is a form of gradient descent derived directly from this soft-margin SVM formulation. For a single training example $(\bx, y)$, let $f(\bx) = \mathbf{w} \cdot \bx + b$. The hinge loss for this example is 
\[
\text{Loss}(\mathbf{w}, b) = \max(0, 1 - y f(\bx)).
\]
The subgradients of this loss with respect to $\mathbf{w}$ and $b$ are:
\begin{equation} \notag
\frac{\partial \text{Loss}}{\partial \mathbf{w}} 
=
\begin{cases}
\lambda \mathbf{w} - y \bx & \text{ if } y f(\bx) < 1, \\[5pt]
\lambda \mathbf{w} & \text{ otherwise},
\end{cases}
\qquad \qquad
\frac{\partial \text{Loss}}{\partial b} 
=
\begin{cases}
- y & \text{ if } y f(\bx) < 1, \\[5pt]
0 & \text{ otherwise}.
\end{cases}
\end{equation}
These subgradients form the basis for our iterative update rules.

\subsection{Iterative Training Algorithm}
\label{sec:iterative_training_algo}

In this section, we give an iterative procedure to train the SVM by adjusting the weight vector $\mathbf{w}$ and the bias $b$ according to an iterative training algorithm.

\medskip

Given data points $\{ (\bx^{(i)}, y^{(i)}) \}^N_{i=1}$, a regularization parameter $\lambda$, and a learning rate $\eta > 0$, the weight $\mathbf{w}$ and the bias $b$ are updated as follows:
\begin{enumerate}
\item Initialize $\mathbf{w} = \mathbf{0}$ and $b = 0$.
    
\item For each data point $(\bx^{(i)}, y^{(i)})$ in the dataset:

\begin{itemize}
\item Compute the decision function: $f(\bx^{(i)}) = \mathbf{w} \cdot \bx^{(i)} + b$.

\item If there is a misclassification or margin violation ($y^{(i)} f(\bx^{(i)}) < 1$), then            
\begin{equation} \label{eq:mis_update}
\begin{split}
\mathbf{w} & \leftarrow \mathbf{w} - \eta (\lambda \mathbf{w} - y^{(i)} \bx^{(i)}) \\
b & \leftarrow b + \eta y^{(i)}
\end{split}
\end{equation}
    
\item If there is no misclassification ($y^{(i)} f(x^{(i)}) \geq 1$), then
\begin{equation} \label{eq:no_mis_update}
\begin{split}
\mathbf{w} & \leftarrow \mathbf{w} - \eta \lambda \mathbf{w} \\
b & \leftarrow b
\end{split}
\end{equation}
\end{itemize}
        
\item Repeat this process for multiple epochs until convergence or a set number of iterations is reached.
\end{enumerate}

This iterative process adjusts $\mathbf{w}$ and $b$ in the direction that minimizes the regularized hinge loss, effectively moving the hyperplane to reduce errors for misclassified points. The learning rate $\eta$ controls the step size of these updates, influencing the convergence rate. In the separable case, these updates converge toward the maximum-margin hyperplane of SVMs. This approach provides a computationally efficient method for training SVMs, particularly beneficial in online or large-scale learning scenarios.

{To implement batch processing in SVM, we modify the update rule. The modification applies specifically to Equation~\eqref{eq:mis_update} and~\eqref{eq:no_mis_update}. The input-dependent parameter updates are aggregated across a batch size $n_{\rm{batch-size}}$, and then we perform a single update. For each sample in the batch, we define
\[
w_{\mathrm{upd}}^{(i)} =
\begin{cases}
\eta y^{(i)} \bx^{(i)}, & \text{if } y^{(i)} f(\bx^{(i)}) < 1,\\
0, & \text{otherwise},
\end{cases}
\qquad
b_{\mathrm{upd}}^{(i)} =
\begin{cases}
\eta y^{(i)}, & \text{if } y^{(i)} f(\bx^{(i)}) < 1,\\
0, & \text{otherwise}.
\end{cases}
\]
The batch update is  given by
\begin{equation} \label{eq:no_mis_update_batch}
\begin{split}
\mathbf{w} &\leftarrow \mathbf{w} - \eta \lambda \mathbf{w} + \frac{1}{n_{\rm{batch-size}}}\sum_{i=1}^{n_{\rm{batch-size}}} w_{\mathrm{upd}}^{(i)}, \\
b &\leftarrow b + \frac{1}{n_{\rm{batch-size}}}\sum_{i=1}^{n_{\rm{batch-size}}} b_{\mathrm{upd}}^{(i)}.
\end{split}
\end{equation}
}

\subsection{Decision Rule}

After training, the classification decision for a new data point $\hat{\bx} \in \mathbb{R}^d$ is determined by
\begin{equation} \notag
    f(\hat{\bx}) = \mathbf{w} \cdot \hat{\bx} + b,
\end{equation}
and the predicted class is given by
\[
\hat{y} = 
\begin{cases} 
1 & \text{ if } f(\hat{\bx}) \geq 0, \\[5pt]
-1 & \text{ otherwise}.
\end{cases}
\]

This iterative approach, while not the standard quadratic programming formulation of SVM, provides a computationally efficient method for training SVMs and is particularly useful in online or large-scale learning scenarios.

\section{Operation Modules}\label{sec:modules}

In this section, we develop the reaction network scheme for the individual submodules that will be used for implementing the \emph{feedforward} and \emph{backpropagation} steps in Sections \ref{sec:feedforward} and \ref{sec:learning}. 
To this end, we note that when dealing with weights and biases, negative values may arise, and a mechanism is required to accommodate them. This issue is addressed in the next subsection on \emph{dual-rail encoding}.

\subsection{Dual-rail Encoding}

In a physical setting, the concentration of species is always non-negative. As data points and weights can be negative, we use the \emph{dual-rail encoding} \cite{vasic2020deep}. 
Specifically, for any variable \( \zeta \in \mathbb{R} \), we take two species \( \zeta^+ \) and \( \zeta^- \) with non-negative concentrations \( \zeta^+(t) \geq 0 \) and \( \zeta^-(t) \geq 0 \). For all $t \geq 0$, the difference between \( \zeta^+(t) - \zeta^-(t) \) represents the real value \( \zeta \).

Here, we list some basic modules from~\cite{vasic2020crn++}. that will be used in the implementation of the feedforward and backpropagation modules. 
We begin by introducing notation that will be used throughout the remainder of this paper.

\medskip

\textbf{Notation: }
We denote by $[X(t)]$ the concentration of species $X$ at time $t$.
If $X$ is a catalyst, whose concentration remains constant for all $t \geq 0$, we write $[X]$.
Finally, $[X^{ss}]$ denotes the concentration of $X$ at steady state.

\subsection{Addition Module}

In order to add two numbers, we consider the following network:
\begin{equation} \label{module_addition}
\begin{split}
A &\overset{1}\longrightarrow A + C \\
B &\overset{1}\longrightarrow B + C \\
C &\overset{1}\longrightarrow \emptyset
\end{split}
\end{equation}

In this network, the original input species $A$ and $B$ are catalysts, and thus their concentrations remain constant. The concentrations of species $A$ and $B$ are added, with the resulting value stored in the steady-state concentration of $C$.
Let $[A] = [A(0)]$ and $[B] = [B(0)]$. 
Then the dynamics associated with the network~\eqref{module_addition} is given by
\begin{equation} \notag
\frac{d[C(t)]}{dt} = [A] + [B] - [C(t)].
\end{equation}
The steady-state concentration of $C$ is given by 
\[
[C^{ss}] = [A] + [B].
\]


\subsection{Multiplication Module}

In order to multiply, we consider the following network:
\begin{equation} \label{module_multiplication}
\begin{split}
A + B &\overset{1}\longrightarrow A + B + C \\
C &\overset{1}\longrightarrow\emptyset
\end{split}
\end{equation}

In this network, the species $A$ and $B$ are catalysts. The concentrations of species $A$ and $B$ are multiplied, with the resulting value stored in the steady-state concentration of $C$.
Let $[A] = [A(0)]$ and $[B] = [B(0)]$. 
Then the dynamics associated with the network~\eqref{module_multiplication} is given by 
\begin{equation} \notag
\frac{d[C(t)]}{dt} = [A] [B] - [C(t)].
\end{equation}
The steady-state concentration of $C$ is given by 
\[
[C^{ss}] = [A] [B].  
\]

\subsubsection{Dual Rail Encoding in Multiplication Module}
We now demonstrate how dual-rail encoding is performed in the multiplication module. Similar mechanisms will hold for other modules in this paper.  

Let $A_p, A_n$ and $B_p, B_n$ be the dual-rail species corresponding to signed quantities $A$ and $B$, respectively, such that
\[
A = [A_p] - [A_n], \qquad B = [B_p] - [B_n].
\]

The objective is to compute the product $C = A \cdot B$ using dual-rail encoding. Towards this, we have
\[
C = A \cdot B =(A_p - A_n)\cdot(B_p - B_n) = A_p B_p + A_n B_n - A_p B_n - A_n B_p = (A_p B_p + A_n B_n) - ( A_p B_n + A_n B_p)
\]

In the dual-rail form, we introduce species $C_p$ and $C_n$ so that
\[
C = [C_p] - [C_n],
\]
where
\[
[C_p] = A_p B_p + A_n B_n, \qquad
[C_n] = A_p B_n + A_n B_p.
\]

We implement $[C_p]$ and $[C_n]$ using the multiplication module \eqref{module_multiplication}. The corresponding reaction network is given by:
\begin{equation} \notag
\begin{split}
A_p + B_p  &\rightarrow A_p + B_p  + C_p \\
A_n + B_n  &\rightarrow A_n + B_n  + C_p \\
A_p + B_n  &\rightarrow A_p + B_n  + C_n \\
A_n + B_p  &\rightarrow A_n + B_p  + C_n \\
C_p  &\rightarrow \emptyset \\
C_n  &\rightarrow \emptyset \\
\end{split}
\end{equation}

The dynamics of the system give 
\[
[C_p^{ss}] = A_p B_p + A_n B_n, \qquad
[C_n^{ss}] = A_p B_n + A_n B_p,
\]
thereby correctly encoding the signed product $C = A \cdot B$ in dual-rail form.

\subsection{Comparison and Approximate Majority Modules}

Our module is designed to compare the concentrations of species and set the corresponding flags. This functionality is achieved using a \textit{comparison module} and an \textit{approximate majority module}, triggered one after another.

\subsubsection*{Comparison Module}

In the comparison module, the inputs $X$ and $Y$ are assigned to flag species $X_{gY}$ and $X_{lY}$ in a normalized fashion. The network for implementing this is given by 
\begin{equation} \label{module_comparison}
\begin{split}
X_{gY} + Y &\overset{1}\longrightarrow X_{lY} + Y \\
X_{lY} + X &\overset{1}\longrightarrow X_{gY} + X
\end{split}
\end{equation}

In this network, the species $X$ and $Y$ are catalysts. Let $[X] = [X(0)]$ and $[Y] = [Y(0)]$. 
To normalize the values within the range $[0, 1]$, we set the initial concentrations of $X_{gY}$ and $X_{lY}$ such that
\[
[X_{gY}(0)] + [X_{lY}(0)] = 1. 
\]
The dynamical system corresponding to the network~\eqref{module_comparison} is given by
\begin{equation} \notag
\begin{split}
\frac{d[X_{gY}(t)]}{dt} &= -[X_{gY}(t)] [Y]  + [X_{lY}(t)] [X], \\
\frac{d[X_{lY}(t)]}{dt} &= [X_{gY}(t)] [Y]  - [X_{lY}(t)] [X].
\end{split}
\end{equation}
Adding the two equations above gives us that $[X_{gY}(t)] + [X_{lY}(t)]$ is preserved for all time $t \geq 0$.
At steady state, we get
\begin{equation} \notag
\frac{[X_{gY}^{ss}]}{[X_{lY}^{ss}]} = \frac{[X]}{[Y]}.
\end{equation}
Since $[X_{lY}^{ss}] + [X_{gY}^{ss}] = [X_{gY}(0)] + [X_{lY}(0)] = 1$, this implies that 
\begin{equation} \notag
[X_{gY}^{ss}] = \frac{[X]}{[X] + [Y]}, \qquad
[X_{lY}^{ss}] =\frac{[Y]}{[X] + [Y]}. 
\end{equation}
Moreover, by direct computation, we can derive that this steady state is globally attracting.
For example, if $[X] = 60$ and $[Y] = 40$, the flag species $X_{gY}$ and $X_{lY}$ will converge to $0.6$ and $0.4$, respectively.

\subsubsection*{Approximate Majority Module} \label{sec:approximate_majority}

We use the following network (\cite{cardelli2012cell}) to implement the approximate majority module:
\begin{equation} \label{module_approximate_majority}
\begin{split}
X_{gY} + X_{lY} &\overset{1}\longrightarrow X_{lY} + B \\
B + X_{lY} &\overset{1}\longrightarrow X_{lY} + X_{lY} \\
X_{lY} + X_{gY} &\overset{1}\longrightarrow X_{gY} + B \\
B + X_{gY} &\overset{1}\longrightarrow X_{gY} + X_{gY}
\end{split}
\end{equation}
where $B$ is an intermediate species. 

The dynamical system corresponding to the network \eqref{module_approximate_majority} is given by
\begin{equation} \notag
\begin{split}
\frac{d[X_{gY}(t)]}{dt} &= -[X_{gY}(t)] [X_{lY}(t)] + [B(t)] [X_{gY}(t)], \\
\frac{d[X_{lY}(t)]}{dt} &= -[X_{gY}(t)] [X_{lY}(t)]   + [B(t)] [X_{lY}(t)], \\
\frac{d[B(t)]}{dt} &= -[B(t)] [X_{lY}(t)]   - [B(t)] [X_{gY}(t)] + 2[X_{gY}(t)][X_{lY}(t)].
\end{split}
\end{equation}
Assuming that the normalization step in the comparison module has already been performed, the species $(X_{gY} (t), X_{lY} (t), B (t))$ will converge to the state $(1, 0, 0)$ if $X_{gY}(0) > X_{lY}(0)$, and to $(0, 1, 0)$ if $X_{gY}(0) < X_{lY}(0)$. 
For a proof of this result, please see \cite{vasic2020crn++}.


\subsection{Loading Module}

The loading module copies the concentration of $A$ into $B$ using the following network:
\begin{equation} \label{module_loading}
\begin{split}
A &\overset{1}{\longrightarrow} A + B \\
B &\overset{1}{\longrightarrow} \emptyset
\end{split}
\end{equation}

In this network, the species $A$ is a catalyst. Let $[A] = [A(0)]$.
The dynamical system corresponding to the network~\eqref{module_loading} is given by 
\begin{equation} \notag
\frac{d[B]}{dt} = [A] - [B].
\end{equation}
At steady state, we get
$[B^{ss}] = [A]$.

\subsection{Subtraction Module}

We define a new species $C$, whose steady-state concentration will store the difference between the concentrations of two species $A$ and $B$. In particular, the subtraction module will compute $[C^{ss}] = \max(0, A - B)$.
The reaction network corresponding to this is given by:
\begin{equation} \label{module_subtraction}
\begin{split}
A &\overset{1}{\longrightarrow} A + C \\
B &\overset{1}{\longrightarrow} B + H \\
C &\overset{1}{\longrightarrow} \emptyset\\
C + H &\overset{1}{\longrightarrow} \emptyset\\
\end{split}
\end{equation}

In this network, the species $A$ and $B$ are catalysts. Let $[A] = [A(0)]$ and $[B] = [B(0)]$. Then the dynamics corresponding to  network~\eqref{module_subtraction} is given by
\begin{equation} \notag
\begin{split}
\frac{d[C(t)]}{dt} &= [A] - [C(t)][H(t)]  - [C(t)], \\
\frac{d[H(t)]}{dt} &= [B]   - [C(t)][H(t)].
\end{split}
\end{equation}
This implies that at steady state,
\begin{equation} \notag
[C^{ss}] =
\begin{cases}
[A] - [B], & \text{ if $[A] > [B]$}, \\[5pt]
0, & \text{ if $[A] \leq [B]$}.
\end{cases}
\end{equation}

\section{Oscillation Module}
\label{sec:hopf_clock}

In the oscillation module, we present a mechanism for time-scale separation using a clock derived from \emph{Hopf dynamics} \cite{kuznetsov2004elements,strogatz2001ndc}.
The Hopf dynamics is turned into a discrete-time signal such that only one group of reactions is activated at a time. (see \textit{Oscillator semantics} at the end of this section for a detailed explanation). 

Note that it is possible to use a standard oscillation module as in~\cite{vasic2020crn++}. However, such oscillations are not guaranteed to maintain constant amplitude and may attenuate over time, leading to errors that accumulate. Consequently, the reaction network implementation may fail to accurately match the implementation of SVMs. To obtain oscillations with constant amplitude, we instead employ a Hopf oscillator, and assume the existence of an external readout that converts this dynamics into a clock signal controlling the sequential execution of reactions (see Remark~\ref{rmk:oscillation} for details).

The Hopf dynamical system \cite{kuznetsov2004elements,strogatz2001ndc} is given by:
\begin{equation}
\label{oscillation_hopf_dynamics}
\begin{split}
\frac{d [X(t)]}{dt} &= \mu [X(t)] - [X(t)]^3 - [Z(t)]^2 [X(t)] - \omega [Z(t)], \\
\frac{d [Z(t)]}{dt} &= \mu [Z(t)] - [Z(t)]^3 - [X(t)]^2 [Z(t)] + \omega [X(t)].
\end{split}
\end{equation}
One can show that the trajectories of this dynamical system converge to a stable limit cycle of radius $\sqrt{\mu}$ and period $T = 2\pi / \omega$. 
We also remark that Equation~\eqref{oscillation_hopf_dynamics} can be realized via mass-action kinetics using dual-rail encoding.

Given the oscillator dynamics, we obtain a discrete-time signal by computing a phase and then discretizing it as follows.
We compute the current clock phase of the oscillator $\theta(t)$ by
\begin{equation} \label{eq:hopf_phase}
\theta(t) = \mathrm{atan2} ([Z(t)], [X(t)]) \bmod 2\pi.
\end{equation}
Here, the function $\mathrm{atan2}(z,x)$ returns the angle between the $x$-axis and the point $(x,z)$.
Next, we discretize the phase into $n_{\texttt{clock}}$ slots as follows:
\begin{equation} \label{eq:phase_to_bin}
k(t) = \left\lfloor \frac{n_{\texttt{clock}}}{2\pi} \theta(t) \right\rfloor \bmod n_{\texttt{clock}},
\end{equation}
where $k(t)$ denotes the active oscillator slot index.
We then define $O(t) \in \{0, 1\}^{n_{\texttt{clock}}}$ by
\[
\text{ For } i \in \{0, \dots, n_{\texttt{clock}} - 1\}, 
\quad
O_i(t) := 
\begin{cases}
1, & \text{ if } i = k(t), \\[5pt]
0, & \text{ otherwise}.
\end{cases}
\]



\begin{figure}[h!]
    \centering    \includegraphics[width=1\linewidth]{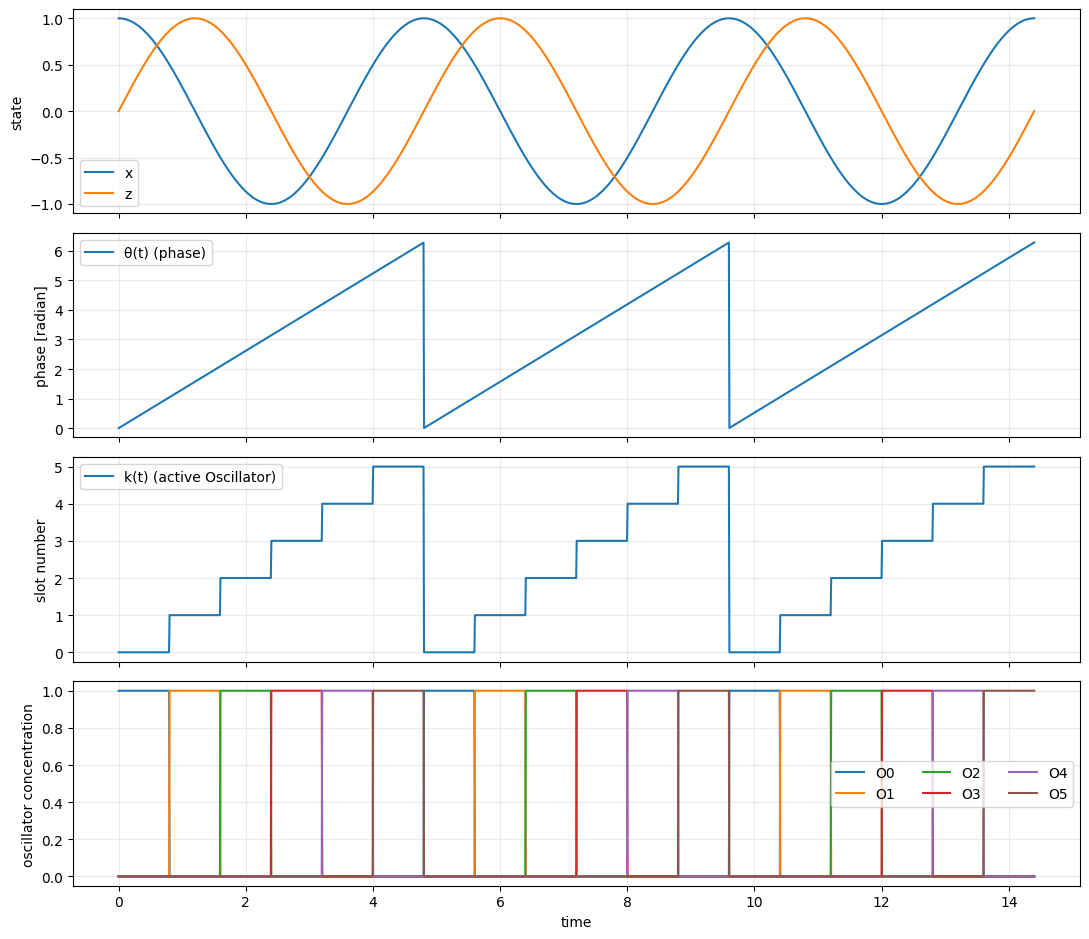}
    \caption{
    Hopf oscillator dynamics for $n_{\texttt{clock}} = 6$. The figure shows the time evolution of (i) $[X(t)]$ and $[Z(t)]$, (ii) the phase $\theta(t)$, (iii) the discrete index $k(t)$, and (iv) the indicator variables $O_i(t)$.
    }
    \label{fig:oscillator}
\end{figure}


\begin{remark}
\label{rmk:oscillation}

The phase computation and discretization in Equations~\eqref{eq:hopf_phase} and~\eqref{eq:phase_to_bin} are not part of the reaction dynamics, but rather constitute a readout operation applied to the observables $x(t)$ and $z(t)$. 
We do not implement this step chemically; instead, we assume the existence of an external measurement device capable of computing this clock signal. This distinction between intrinsic dynamics and external readouts is standard in bioengineering applications.
\end{remark}

\paragraph{Oscillator semantics:}

Every reaction in our SVM implementation is associated with an oscillator index $i \in \{0, \dots, n_{\texttt{clock}} - 1\}$. If the intrinsic rate of a reaction $r$ is $a_r(\cdot)$, then its effective propensity in the oscillator molecules is given by
\begin{equation} \notag
a_r^{\texttt{eff}}(\cdot, t) = O_i(t)\, a_r(\cdot)
\ \text{ with } 
i \in \{0, \dots, n_{\texttt{clock}} - 1\}.
\end{equation}
The oscillator species \( O_i \) acts as a binary gate for its associated reactions. At time $t$, when $O_i(t) = 1$, the corresponding reactions are active and proceed at rates, i.e., $a_r^{\texttt{eff}}(\cdot, t) = a_r(\cdot)$. When $O_i(t) = 0$, the effective rate is zero and the reactions are inactive. In this way, the oscillator enforces a sequential execution of reactions according to the clock signal.
Moreover, the oscillator molecule functions as a catalyst: it modulates the reaction rates without being  altered by the reactions.

\section{Assignment Module}
\label{sec:assignment}

Inspired by~\cite{fan2023automatic}, we develop the assignment module, which divides the input samples into batches and loads them into the input species set. This strategy allows us to parallelize our computation and allows faster execution of our program.

To formulate this, assume we have $p$ samples, each of dimension $d$. 
In each iteration, we aim to feed $\tilde{p}$ samples (typically $\tilde{p}$ divides $p$) into the input layer.
We divide the $p$ samples across $\tilde{p}$ parallel input lanes. For every lane $l \in \{1,\ldots,\tilde{p}\}$, the eligible sample indices form the  set
\[
\mathcal{I}_l := \{l,\tilde{p}+l,2\tilde{p}+l,\ldots\, p-\tilde{p}+l \}.
\]
In every iteration, a sample is chosen from every lane. These samples are fed into the input layer. Over successive iterations, the active samples in each $\mathcal{I}_l$ are shifted through to be chosen. After the last sample is chosen, we cycle back to the first sample and continue the process.

For simplicity, we assume that all sample values are non-negative. 
We define
\begin{itemize}
\item The sample species set: $\{I_1^i, I_2^i, I_3^i\}$, where $1 \leq i \leq p$, represents the set of input samples.
\item The input species set: $\{X_1^l, X_2^l, Y^l\}_{l=1}^{\tilde{p}}$, where $1 \leq l \leq \tilde{p}$, represents the set of input samples fed into the network at a time.

\item The order species set: $\{C_i^l\}$, where $1 \leq i \leq p$ and $1 \leq l \leq \tilde{p}$.

\item The auxiliary order species set: $\{\tilde{C}_i^l\}$, where $1 \leq i \leq p$ and $1 \leq l \leq \tilde{p}$.
\end{itemize}

These species support the assignment module, which operates in three phases. 
The reactions are
\begin{equation} \notag
\begin{split}
O_0: \quad 
& C_i^l + I_q^i \xrightarrow{1} C_i^l + I_q^i + X_q^l \qquad X_q^l \xrightarrow{1} \emptyset \\
& C_i^l + I_3^i \xrightarrow{1} C_i^l + I_3^i + Y^l \qquad Y^l \xrightarrow{1} \emptyset  \\
O_1: \quad 
& C_i^l \xrightarrow{k} \tilde{C}_i^l  \\
O_2: \quad 
& \tilde{C}_o^l \xrightarrow{k} C_{\tilde{p} + o}^l \qquad 
\tilde{C}_{p - \tilde{p} + l}^l \xrightarrow{k} C_l^l \qquad 
\tilde{C}_j^l \xrightarrow{k} C_j^l 
\end{split}
\end{equation}
where $q \in \{1,2\}$, $o \in \mathcal{I^{``}}_l$, with $\mathcal{I^{``}}_l := \{ l, \tilde{p} + l, 2\tilde{p} + l, \ldots, p - 2\tilde{p} + l\}$, and $j \notin \mathcal{I}_l.$ 

To initialize the system, we set the initial concentrations of the order species $\{C_i^l\}$ as
\[
[C_i^l(0)] = 
\begin{cases}
1, & \text{ if } l = i, \\[5pt]
0, & \text{ otherwise},
\end{cases}
\ \text{ for every $1 \leq i \leq p$ and $1 \leq l \leq \tilde{p}$}.
\]
 
This means that for each lane index $l \in \{1, \ldots,\tilde{p}\}$, only $C_l^l$ are active initially, while all other order species remain inactive. For the auxiliary order species $\{\tilde{C}_i^l\}$, all concentrations are initialized to zero as follows:
\[
[\tilde{C}_i^l(0)] = 0 
\ \text{ for every $1 \leq i \leq p$ and $1 \leq l \leq \tilde{p}$}.
\]
The concentrations of the input species  $\{Y^l\}$ and  $\{X_q^l\}$ (for $q \in \{1,2\}$ and $l \in \{1, \ldots, \tilde{p}\}$) are initialized to arbitrary nonnegative values:
\[ 
[Y^l(0)] \cup [X_q^l(0)] \in \mathbb{R}_{\geq 0}
\ \text{ for every $1 \leq q \leq 3$ and $1 \leq l \leq \tilde{p}$}.
\]

In Reaction Phase $O_0$, the assignment operation is performed. Specifically, the reactions:
\begin{equation} \notag
\begin{split}
C_i^l + I_q^i & \xrightarrow{1} C_i^l + I_q^i + X_q^l
\\
C_i^l + I_3^i & \xrightarrow{1} C_i^l + I_3^i + Y^l \\
X_q^l & \xrightarrow{1} \emptyset \\
Y^l & \xrightarrow{1} \emptyset
\end{split}
\end{equation}
This reaction network collectively stores the sum of the term-wise product of the order species concentration $[C_i^l]$ with the sample input species $I_1^i$, $I_2^i$ and $I_3^i$ into the corresponding $X_1^l$, $X_2^l$ and $Y^l$. Since the catalyst concentrations satisfy $[C_i^l] = 0$ for all $l \ne i$ and $[C_l^l] = 1$, the system effectively selects and loads only one active sample per lane during each iteration. Thus, $\tilde{p}$ samples are loaded into the input layer parallel.

In the Phase $O_1$, the active order species are copied into the auxiliary species via the reaction:
\[
C_i^l \xrightarrow{k} \tilde{C}_i^l
\]
This mechanism preserves the current assignment state in preparation for the rotation of order species in the next iteration.

In Reaction Phase $O_2$, the system initiates a cyclic shift of the active  samples across lanes. The reactions:
\[
\tilde{C}_o^l \xrightarrow{k} C_{\tilde{p}+o}^l \qquad
\tilde{C}_{p - \tilde{p} + l}^l \xrightarrow{k} C_l^l \qquad
\tilde{C}_j^l \xrightarrow{k} C_j^l
\]
perform the following: the active block in each position is shifted forward by $\tilde{p}$, going to the next element in their respective $\mathcal{I}_l$, except for the last element of the lane, which is cyclically wrapped around to the first position. Any remaining auxiliary species that do not participate in the cycle are transferred to their corresponding order species. 

\begin{remark}
    
As per the above convention, reactions are formed for all pairs of $(l, i)$. However, for a given $l$, only those $i \in \mathcal{I}_l$ have $C_i^l$ as active at some time during the network.  Consequently, reactions involving those inactive indices will never contribute to the dynamics.
In our Python implementation, we iterate only over those indices that can be active and skip inactive indices to reduce computational overhead.

Together, these phases constitute an automated mechanism for batch-wise data assignment, ensuring that $\tilde{p}$ samples are loaded in parallel at each iteration and that the active sample in every lane evolves cyclically without external logic.
\end{remark}

\section{Encoding the Feedforward Module of SVM}\label{sec:feedforward}

In the following sections, we describe the computation done for a single sample passing through the reaction network. We omit the lane indices used previously. The weights and biases are shared across lanes. But the input, and intermediate species are lane specific, {unless stated otherwise}.  

Given an input sample $\bx = (x_1, \ldots, x_d) \in \mathbb{R}^d$ with an associated label $y \in \{ -1, 1 \}$, a weight vector $\mathbf{w} = (w_1, \ldots, w_d) \in \mathbb{R}^d$, and a bias term $b \in \mathbb{R}$, our objective is to determine whether the classification is correct by evaluating $y(\mathbf{w}\cdot \bx + b)$ against $1$.

In this section, we construct reaction networks that compute the products $w_i x_i$, sum them together with the bias $b$, and then multiply the resulting sum by $y$ (Steps 1–3), allowing the resulting value to be compared with 1 (Steps 4–5).
The complete computation is performed through a sequence of five steps, utilizing the oscillatory molecules $O_3$, $O_4$, $O_5$, $O_6$, $O_7$ and $O_8$.

In Sections \ref{sec:feedforward} and \ref{sec:learning}, we denote by $W_i$ the species representing the weight $w_i$, by $X_i$ the species representing $x_i$, by $Y$ the species representing $y$, and by $B$ representing the bias $b$. 

\textbf{Step 1:}
We introduce the species $WX_i$, whose steady state concentration will equal the product of $W_i$ and $X_i$. Specifically, we use the multiplication module \eqref{module_multiplication} and consider the following network:
\begin{equation}\label{eq:W_dot_X}
\begin{split}
W_1 + X_1 + O_3  &\overset{1}\longrightarrow W_1 + X_1 + WX_1 + O_3 \\
WX_1 + O_3 &\overset{1}\longrightarrow O_3 \\
\vdots & \\
W_d + X_d + O_3  &\overset{1}\longrightarrow W_d + X_d + WX_d + O_3 \\
WX_d + O_3 &\overset{1}\longrightarrow O_3 
\end{split}
\end{equation}

In this network, the species $W_i$, $X_i$ are catalysts. Let $[W_i] = [W_i(0)]$ and $[X_i] = [X_i(0)]$. The dynamical system corresponding to the network \eqref{eq:W_dot_X} is given by
\begin{equation} \notag
\frac{d[WX_i (t)]}{dt} = [W_i] [X_i] [O_3] - [WX_i (t)] [O_3]
\ \text{ for every $1 \leq i \leq d$}.
\end{equation}
At steady state, we get $[{WX_i}^{ss}] = [W_i][X_i]$.

\textbf{Step 2:}
We introduce a new species $P$, whose steady-state concentration represents the sum of all $WX_i$ plus $B$. In particular, we use the addition module \eqref{module_addition} and consider the following network:
\begin{equation}\label{eq:step_2}
\begin{split}
WX_1 + O_4 &\overset{1}\longrightarrow WX_1 + P + O_4 \\
\vdots & \\
WX_d + O_4 &\overset{1}\longrightarrow WX_d + P + O_4 \\
B + O_4 &\overset{1}\longrightarrow B + P + O_4 \\
P + O_4 &\overset{1}\longrightarrow O_4
\end{split}
\end{equation}

In this network, the species $WX_i, B$ are catalysts. Let $[WX_i] = [WX_i(0)]$ and $[B] = [B (0)]$.
The dynamical system corresponding to the network \eqref{eq:step_2} is given by
\begin{equation} \notag
\frac{d[P (t)]}{dt} = \displaystyle\sum_{i=1}^d [WX_i] [O_4]  + [B][O_4] - [P (t)][O_4].
\end{equation}
At steady state, we get
$[P^{ss}] = \displaystyle\sum_{i=1}^d [WX_i] + [B]$.

\textbf{Step 3:}
We introduce the species $Q$, whose steady-state concentration corresponds to the product of $P$ and $Y$.
Specifically, we use the multiplication module \eqref{module_multiplication} and consider the following network:
\begin{equation}\label{eq:step_3}
\begin{split}
P + Y + O_5 &\overset{1}\longrightarrow P + Y + Q + O_5 \\
Q + O_5 &\overset{1}\longrightarrow O_5
\end{split}
\end{equation}

In this network, the species $P$, $Y$ are catalysts. Let $[P] = [P (0)]$ and $[Y] = [Y (0)]$. The dynamical system corresponding to network~\eqref{eq:step_3} is 
\begin{equation} \notag
\frac{d[Q (t)]}{dt} = [P][Y][O_5] - [Q (t)][O_5].
\end{equation}
At steady state, we get $[Q^{ss}] = [P] [Y]$. Combining this with the results from Steps 1 and 2, it follows that
\begin{equation} \notag 
[Q^{ss}] = \Big( \displaystyle\sum_{i=1}^N [WX_i] + [B] \Big) [Y]. 
\end{equation}

Next, we compare the value stored in $Q$ with $1$. 
Depending on whether the associated label is correct (i.e., whether the concentration of $Q$ is greater or less than $1$), the concentration of one of the two flag species ($K_{lQ}$ and $K_{gQ}$) converges to $1$, while the concentration of the other species converges to $0$.

We divide this process into two sequential steps: Step 4, which normalizes the flag species, and Step 5, which performs the comparison. To ensure these steps occur sequentially, we employ oscillatory molecules. Oscillation molecule $O_7$ is used in Step $4$, while $O_{8}$ is used in Step $5$. This guarantees that the reactions in Step 4 complete before those in Step 5 begin.

We employ dual-rail encoding to ensure that all species concentrations remain non-negative. Since the quantity $Q$ can take negative values, we compute it using the subtraction module (Equation \eqref{module_subtraction} ), which produces an effective concentration proportional to $\max(Q^+ - Q^-, 0)$. This introduces a rectification that clamps negative values of $Q$ to zero. However, this does not affect correctness, as the subsequent comparison only depends on whether $Q$ exceeds the threshold; if $Q$ were negative, clamping it to zero preserves the outcome of the comparison. We utilise $O_6$ molecule in this step.

\textbf{Step 4:}
We introduce a catalytic species $K$ along with the flag species $K_{lQ}$ and $K_{gQ}$. 
We set the concentration of $K$ to $1$, i.e., $[K] = [K(0)] = 1$, and designate $K_{lQ}$ and $K_{gQ}$ as flag species, with the sum of their initial concentrations equal to $1$.
Using the comparison module \eqref{module_comparison}, the network implementing this comparison is given by 
\begin{equation}\label{eq:step_4}
\begin{split}
Q + K_{gQ} + O_7 &\overset{1}\longrightarrow K_{lQ} + Q + O_7 \\
K_{lQ} + K + O_7 &\overset{1}\longrightarrow K_{gQ} + K + O_7 
\end{split}
\end{equation}

In this network, the species $Q$ is a catalyst. Let $[Q] = [Q(0)]$ and let $[K_{lQ}(0)] + [K_{gQ}(0)]  = 1$. The dynamical system corresponding to network~\eqref{eq:step_4} is given by 
\begin{equation} \notag
\begin{split}
\frac{d[K_{lQ} (t)]}{dt} &= [Q][K_{gQ} (t)][O_7] - [K_{lQ} (t)][K][O_7], \\
\frac{d[K_{gQ} (t)]}{dt} &= -[Q][K_{gQ} (t)][O_7] + [K_{lQ} (t)][K][O_7]. 
\end{split}
\end{equation}
Adding the two equations above gives us that $[K_{lQ}(t)] + [K_{gQ}(t)]$ is preserved for all time $t \geq 0$.
Consequently, the steady-state concentrations of $K_{lQ}$ and $K_{gQ}$ will be set as follows:
\begin{equation} \notag
\begin{split}
[K_{lQ}^{ss}] = \frac{[Q]}{[Q]+[K]} = \frac{[Q]}{[Q]+1}, \qquad
[K_{gQ}^{ss}] = \frac{[K]}{[Q]+[K]} = \frac{1}{[Q]+1}.
\end{split}
\end{equation}

\textbf{Step 5:}
We compare $Q$ and $K$ by using the approximate majority algorithm (see Section~\ref{sec:approximate_majority}).
Specifically, the flag species $K_{lQ}$ and $K_{gQ}$ are set such that, if $[Q] > [K] = 1$, $K_{lQ}$ converges to $1$ and $K_{gQ}$ converges to $0$; conversely, if $[Q] < [K] = 1$, $K_{lQ}$ converges to $0$ and $K_{gQ}$ converges to $1$.
Using the approximate majority module \eqref{module_approximate_majority}, the network that implements this behavior is given by
\begin{equation}\label{eq:step_5}
\begin{split}
K_{gQ} + K_{lQ} + O_8 &\overset{1}\longrightarrow K_{lQ} + U + O_8 \\
U + K_{lQ} + O_8 &\overset{1}\longrightarrow 2 K_{lQ} + O_8 \\
K_{lQ} + K_{gQ} + O_8 &\overset{1}\longrightarrow K_{gQ} + U + O_8 \\
U + K_{gQ} + O_8 &\overset{1}\longrightarrow 2 K_{gQ} + O_8 
\end{split}
\end{equation}

The dynamical system corresponding to network~\eqref{eq:step_5} is given by
\begin{equation} \notag
\begin{split}
\frac{d[K_{lQ} (t)]}{dt} & = -[K_{gQ} (t)][K_{lQ} (t)][O_8] + [U (t)][K_{lQ} (t)][O_8], \\
\frac{d[K_{gQ} (t)]}{dt} & = -[K_{gQ} (t)][K_{lQ} (t)][O_8] + [U (t)][K_{gQ} (t)][O_8], \\
\frac{d[U (t)]}{dt} & = [K_{gQ} (t)][K_{lQ} (t)][O_8] - [U (t)][K_{gQ} (t)][O_8] + [K_{lQ} (t)][K_{gQ} (t)][O_8] \\ & \qquad - [U (t)][K_{gQ} (t)][O_8].
\end{split}
\end{equation}

The analysis of this dynamical system follows the discussion in Section~\ref{sec:approximate_majority}. In conclusion, if $K_{lQ} (t)$ converges to 1 and $K_{gQ}(t)$ converges to 0, this indicates that $[Q] > [K]$ and the associated label is correct. Conversely, if $K_{lQ} (t)$ converges to 0 and $K_{gQ}(t)$ converges to 1, this indicates that $[Q] < [K]$ and the associated label is incorrect.

\section{Backpropagation Module of SVM using Gradient Descent}\label{sec:learning}

In this section, we construct reaction networks corresponding to the backpropagation component of the SVM, that is, the adjustment of the weight vector $\mathbf{w}$ and the bias $b$ via gradient descent.

Recall from Section~\ref{sec:feedforward} that we determined whether a given input sample $\bx$ with label $y$ is correctly classified under $\mathbf{w}$ and $b$.
Given a regularization parameter $\lambda$ and a learning rate $\eta > 0$, the reaction networks implemented here realize Equations~\eqref{eq:mis_update} and~\eqref{eq:no_mis_update} from the iterative training algorithm, given by:
\begin{itemize}
\item Misclassification case ($y (\mathbf{w} \cdot \bx + b) < 1$):            
\begin{equation} \notag
\begin{split}
\mathbf{w} & \leftarrow \mathbf{w} - \eta (\lambda \mathbf{w} - y \bx) \\
b & \leftarrow b + \eta y
\end{split}
\end{equation}
    
\item No misclassification case ($y (\mathbf{w} \cdot \bx + b ) \geq 1$):
\begin{equation} \notag
\begin{split}
\mathbf{w} & \leftarrow \mathbf{w} - \eta \lambda \mathbf{w} \\
b & \leftarrow b
\end{split}
\end{equation}
\end{itemize}

We first compute the scaled weight $(1-\eta\lambda)\mathbf{w}$ (Step 1), since this term appears in both cases regardless of classification outcome. Next, we compute the products $\eta y$ and $\eta y \bx$, which are required only in the misclassification case (Step 2). Finally, we apply the updates to the bias and the weight vector (Step 3).
The complete computation is performed in two phases, utilizing the oscillatory molecules $O_{9}$ and $O_{10}$.



\textbf{Step 1:}
We introduce the species $R_i$, whose steady state concentration will equal the product of $W_i$ and $\alpha= 1-\eta \lambda$ (update for the weight).
Specifically, we we use the multiplication module \eqref{module_multiplication} and consider the following network:
\begin{equation} \label{eq:W_update_1}
\begin{split}
O_9 + \alpha  + W_1 &\rightarrow O_9 + \eta + \lambda\alpha + W_1 + R_1 \\
R_1 + O_9 &\rightarrow O_9\\
\vdots &\\
O_9 + \alpha + W_d &\rightarrow O_9 + \eta+\lambda\alpha + W_d + R_d \\
R_d + O_9 &\rightarrow O_9\\
\end{split}
\end{equation}

In this network, the species $W_i$ is a catalyst. Let $[W_i] = [W_i(0)]$. The dynamical system corresponding to the network \eqref{eq:W_update_1} is given by
\begin{equation} \notag
\frac{d[R_i (t)]}{dt} = {\alpha} [W_i] [O_9] - [R_i (t)][O_9].
\end{equation}
At steady state, we get $[R_i^{ss}] = {\alpha} [W_i] = (1-\eta\lambda)[W_i]$.

Next, we load the bias. We introduce the species $Z$, whose steady-state concentration $B$. 
Using the loading module \eqref{module_loading}, the reaction network implementing this operation is given by:

\begin{equation} \notag
\begin{split}
B + O_{9} &\rightarrow O_{9} + B + Z \\
Z +  O_{9} &\rightarrow O_{9} \\
\end{split}
\end{equation}

In this network, the species $B$ is catalyst. Let $[B] = [B(0)]$. The dynamical system corresponding to this network is
\begin{equation} \notag
\frac{d[Z (t)]}{dt} = [B] [O_{9}] - [Z (t)] [O_{9}].
\end{equation}
At steady state, we get $[Z^{ss}] = [B]$.

{Note that both the species, $R_i$ and $Z$ are common to all the lanes.}

\textbf{Step 2:}
We introduce two species $XY_i$ and {$Sb$}, whose steady-state concentrations represent the quantities $\eta K_{gQ} X_i Y$ (update for the weight) and $\eta K_{gQ} Y$ (update for the bias), respectively. In particular, we use the multiplication module \eqref{module_multiplication} and consider the following network:
\begin{equation} \notag
\begin{split}
O_9 + \eta + Y + X_1 + K_{gQ} &\rightarrow O_9 + \eta + Y + X_1 + XY_1 + K_{gQ} \\
XY_1 + O_9 &\rightarrow O_9 \\
\vdots & \\
O_9 + \eta + Y + X_d + K_{gQ} &\rightarrow O_9 + \eta + Y + X_d + XY_d + K_{gQ} \\
XY_d + O_9 &\rightarrow O_9 \\
O_9 + \eta + Y  + K_{gQ} &\rightarrow O_9 + \eta + Y + Sb + K_{gQ} \\
Sb + O_9 &\rightarrow O_9 \\
\end{split}
\end{equation}

In this network, the species $X_i, Y, K_{gQ}$ are catalysts. Let $[X_i] = [X_i(0)]$, $[Y] = [Y(0)]$, and $[K_{gQ}] = K_{gQ} (0)$. The dynamical system corresponding to this network is
\begin{equation} \notag
\begin{split}
\frac{d[XY_i (t)]}{dt} & = \eta [K_{gQ} (t)] [X_i] [Y] [O_9] - [XY_i (t)][O_9], \\
\frac{d[Sb (t)]}{dt} & = \eta [K_{gQ} (t)]  [Y] [O_9] - [Sb (t)][O_9].
\end{split}
\end{equation}
At steady state, we get $[XY_i^{ss}] = \eta [K_{gQ}] [X_i] [Y]$ and $[Sb^{ss}] = \eta [K_{gQ}] [Y]$.
Note that the catalyst species $K_{gQ}$ acts as a switch: it takes the value $[K_{gQ}] = 1$ in the misclassification case and $[K_{gQ}] = 0$ in the no-misclassification case, thereby enabling or preventing weight updates.

Note that the species $XY_i$ and $Sb$ are shared across lanes, but each lane contributes to their production independently. In particular, each lane computes its local update term, and these contributions are accumulated into the shared species through reactions whose rate constants are scaled by $1/\tilde{p}$. At steady state, the accumulated quantities correspond to the mean over the $\tilde{p}$ samples processed in parallel. Thus,

$[XY_i^{ss}] = \frac{1}{\tilde{p}} \sum_{l=1}^{\tilde{p}} {\eta }[K_{gQ}^{(l)}][X_i^{(l)}][Y^{(l)}]$, \quad
$[Sb^{ss}] = \frac{1}{\tilde{p}} \sum_{l=1}^{\tilde{p}} \eta [K_{gQ}^{(l)}][Y^{(l)}]$.

\textbf{Step 3:}
We now update the weight and the bias.

{First, we recall from Step 1 that the species $R_i$ directly encodes the scaled weight $(1-\eta\lambda)W_i$. We directly use $R_i$ as the base weight and add the correction term $XY_i$ corresponding to the hinge-loss update.}
We directly combine $R_i$ and the correction term $XY_i$ to form the updated weight. To update the bias, we store the sum of $Z$ and $Sb$ into $B$
The reaction network implementing this operation is given by:

{
\begin{equation}
\begin{split}
R_i + O_{10} &\rightarrow R_i + O_{10} + W_i \\
XY_i + O_{10} &\rightarrow XY_i + O_{10} + W_i \\
Z + O_{10} &\rightarrow Z + O_{10} + B \\
Sb + O_{10} &\rightarrow Sb + O_{10} + B \\
W_i + O_{10} &\rightarrow O_{10} \\
B + O_{10} &\rightarrow O_{10} \\
\end{split}
\end{equation}
}

{In this network, the species $R_i$, $XY_i$, $Sb$ and $Z$ are catalysts. Let $[R_i] = [R_i (0)]$, $[XY_i] = [XY_i (0)]$, $[Sb] = [Sb(0)]$ and $[Z] = [Z (0)]$. The dynamical system corresponding to this network is}
\begin{equation} \notag
\begin{split}
\frac{d [W_i (t)]}{dt} = [R_i][O_{10}] + [XY_i][O_{10}] - [W_i (t)][O_{10}], \\
\frac{d [B (t)]}{dt} = [Z][O_{10}] + [Sb][O_{10}] - [B (t)][O_{10}].
\end{split}
\end{equation}

At steady state, we obtain $[W_i^{ss}] = [R_i] + [XY_i]$ and $[B^{ss}] = [Z] + [Sb]$.
{As a result, Equations~\eqref{eq:mis_update} and~\eqref{eq:no_mis_update} from the iterative training algorithm are implemented (Equation \eqref{eq:no_mis_update_batch} in our parallel batch setting).  After the update is applied, all intermediate species are reset in $O_{11}$ {by using decay reactions}, restoring the system to its initial state for the next iteration. This process is repeated across all batches until a predefined number of iterations is reached.}

\section{Results and Analysis}\label{sec:analysis}

In this section, we compare the actual and predicted values of the learned SVM parameters across the {reaction network}.


{\subsection{Experimental Setup}}

{To validate the functionality of our reaction network, we perform simulations using a CRN-based SVM and compare its performance against a standard soft-margin SVM baseline,  implemented in Python. We generate a two-dimensional linearly separable dataset using a random generator and introduce controlled noise. Each sample is assigned a binary label in $\{+1, -1\}$.}

In our experimental setup, we consider $p = 10$ training samples (evenly divided between the two classes) and 4 test samples. We use a minibatch size of $\tilde{p} = 2$ and train the model for $100$ epochs.
\subsection{Feedforward Network and Learning Module Simulation Results}

\begin{table}[!h]
    \centering
    \begin{tabular}{lccc}
        \toprule
        \textbf{Parameter} & \textbf{Actual Value} & \textbf{Net Predicted Value} & \textbf{Absolute Error} \\
        \midrule
        $w_1$ & 1.2548 & 1.2548 & 0.0 \\
        $w_2$ & -0.2598 &-0.2598 & 0.0 \\
        $b$   & -0.1 & -0.0999 & 0.0001 \\
        \midrule
        \textbf{Average Absolute Error} & & & \textbf{0.0001} \\
        \bottomrule
    \end{tabular}
    \caption{Comparison of Actual and Predicted SVM Weights}
    \label{tab:svm_weights}
\end{table}

We compare the actual and predicted values of the learned SVM parameters. The table below presents a comparison of the true and computed values for weights and bias, along with their respective errors.


The results indicate that the average error in predicting the parameters is approximately  \textbf{{$0.033\%$}}, demonstrating that the CRN-based approach closely approximates the standard SVM training process.


\begin{figure}[!h]
    \centering
    \begin{subfigure}{0.43\textwidth}
        \centering
        \includegraphics[width=\linewidth]{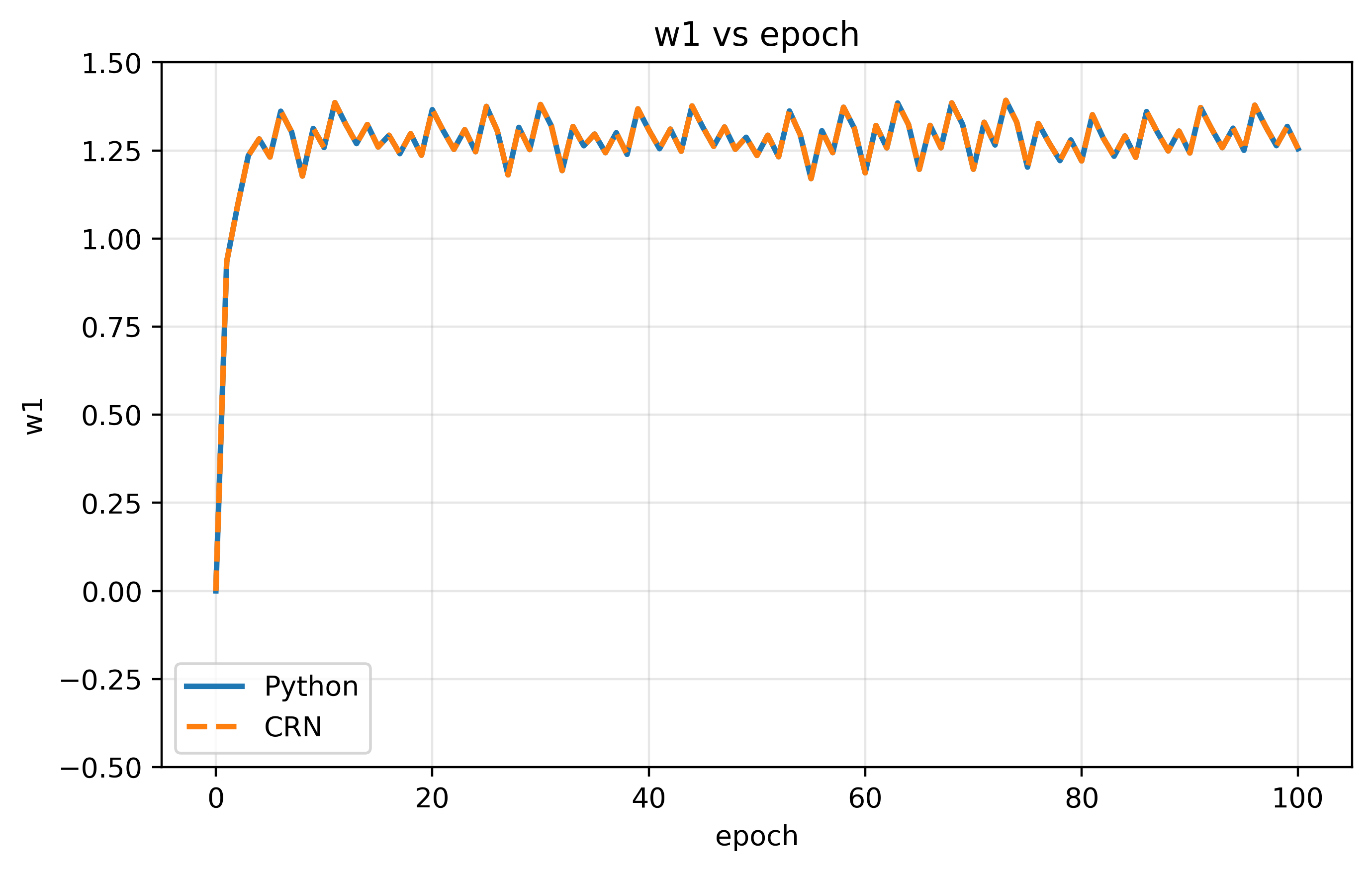}
        \caption{ Weight 1 as a function of epochs}
        \label{fig:w1}
    \end{subfigure}
    \hfill
    \begin{subfigure}{0.43\textwidth}
        \centering
        \includegraphics[width=\linewidth]{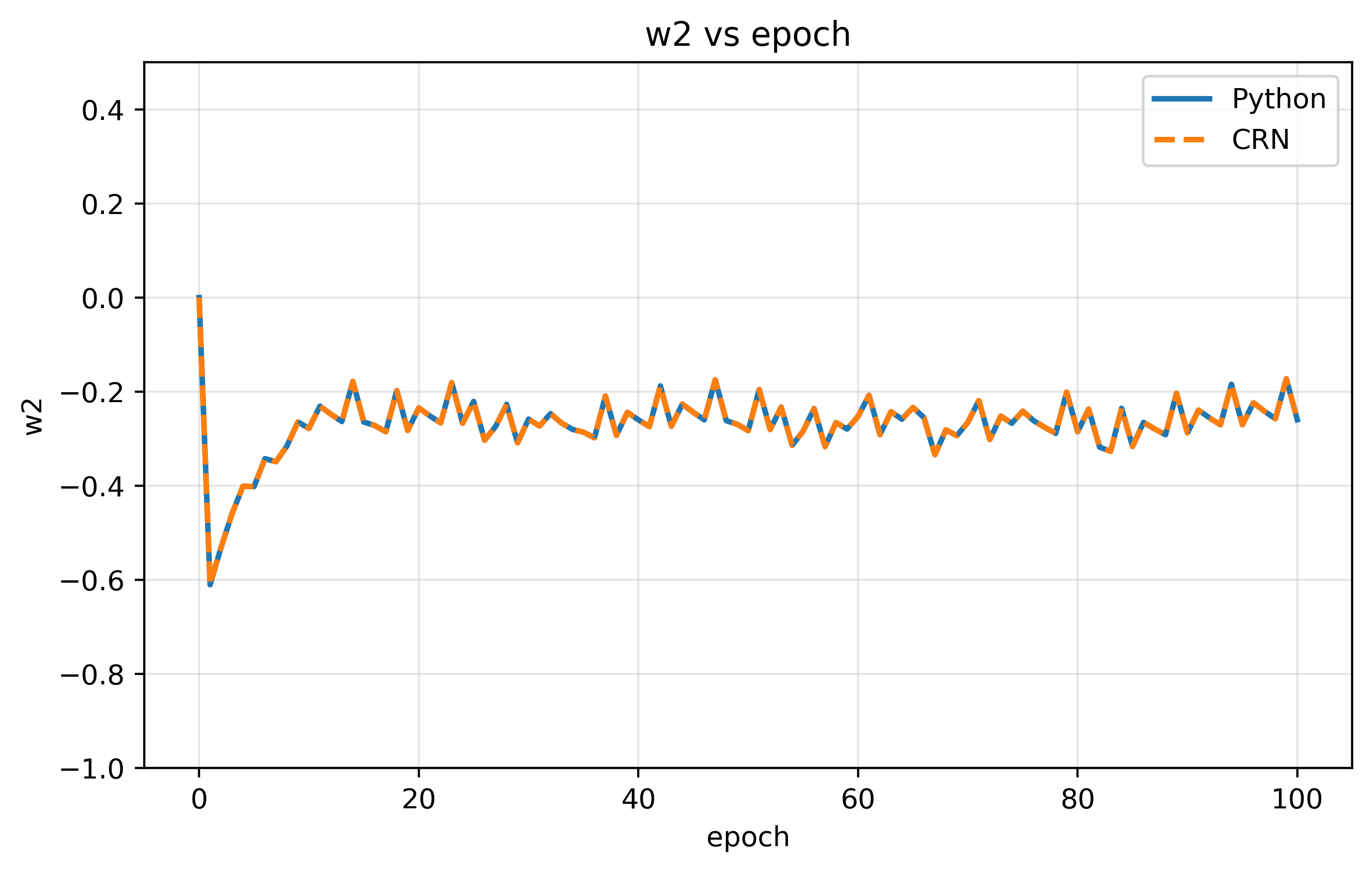}
        \caption{ Weight 2 as a function of epochs}
        \label{fig:w2}
    \end{subfigure}
    
    \vspace{0.2cm}
    
    \begin{subfigure}{0.43\textwidth}
        \centering
        \includegraphics[width=\linewidth]{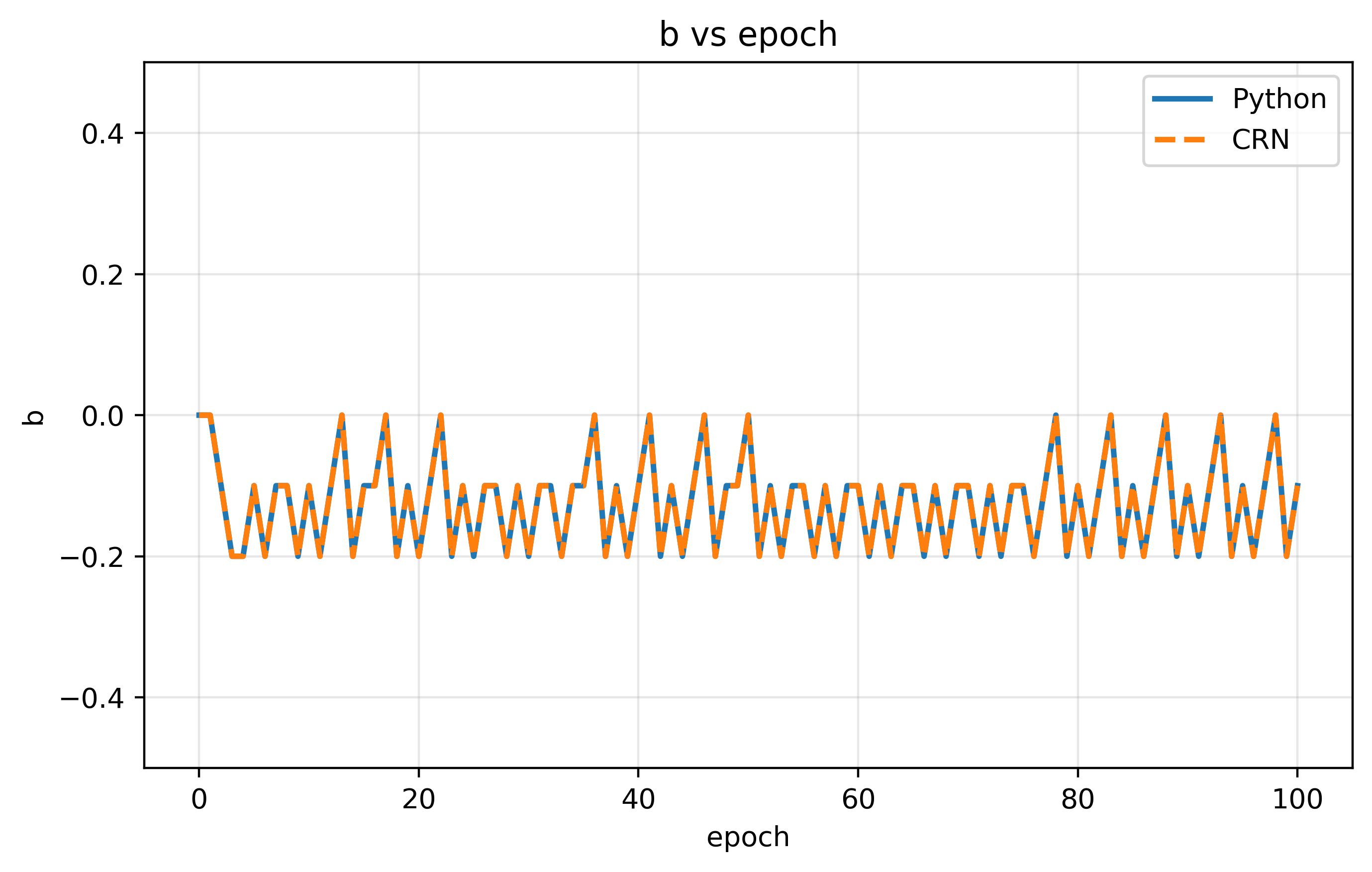}
        \caption{Bias as a function of epochs}
        \label{fig:bias}
    \end{subfigure}
    \hfill
    \begin{subfigure}{0.43\textwidth}
        \centering
        \includegraphics[width=\linewidth]{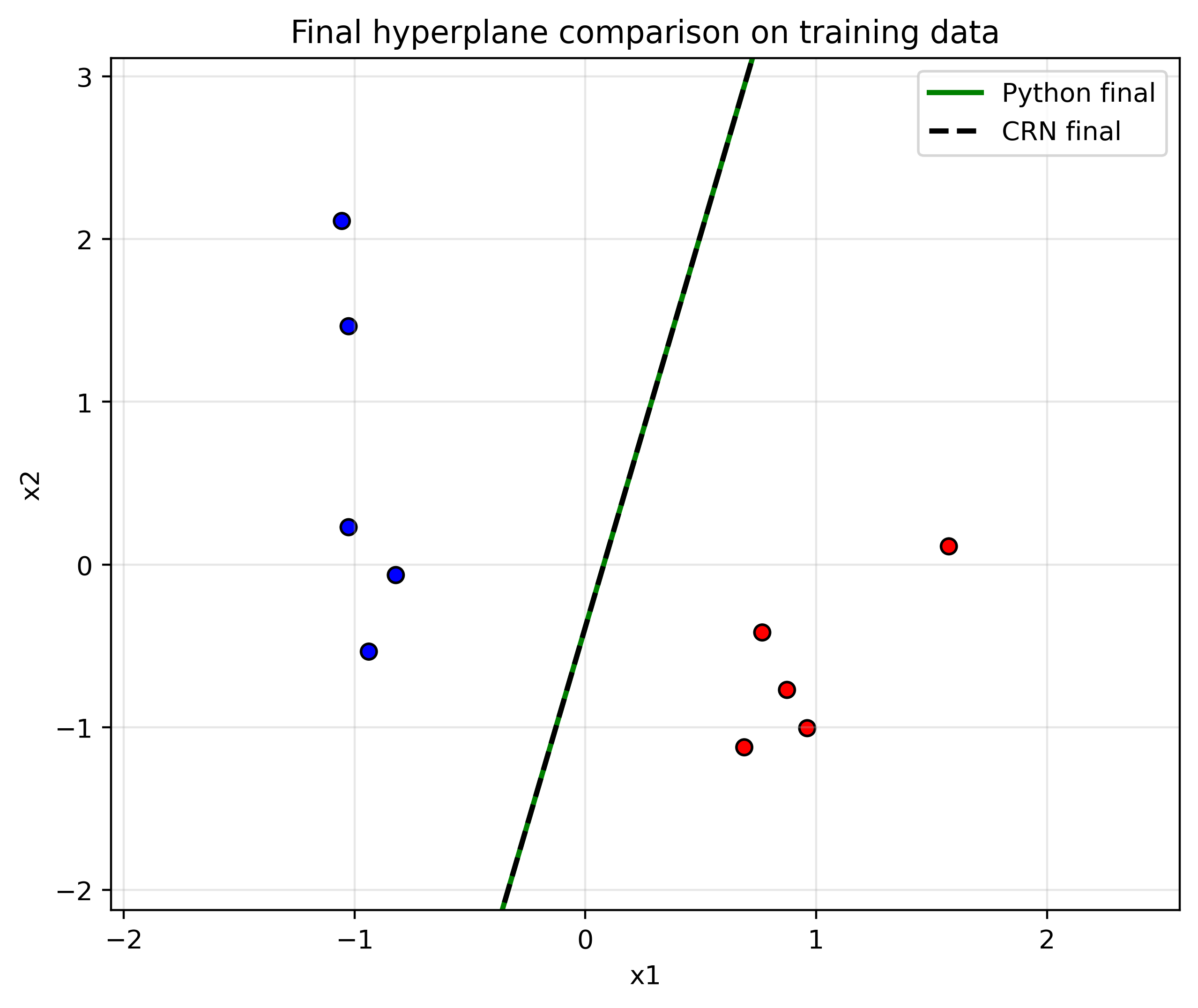}
        \caption{Hyperplane on Train Data}
        \label{fig:train_hyperplane}
    \end{subfigure}
    
    \begin{subfigure}{0.43\textwidth}
        \centering
        \includegraphics[width=\linewidth]{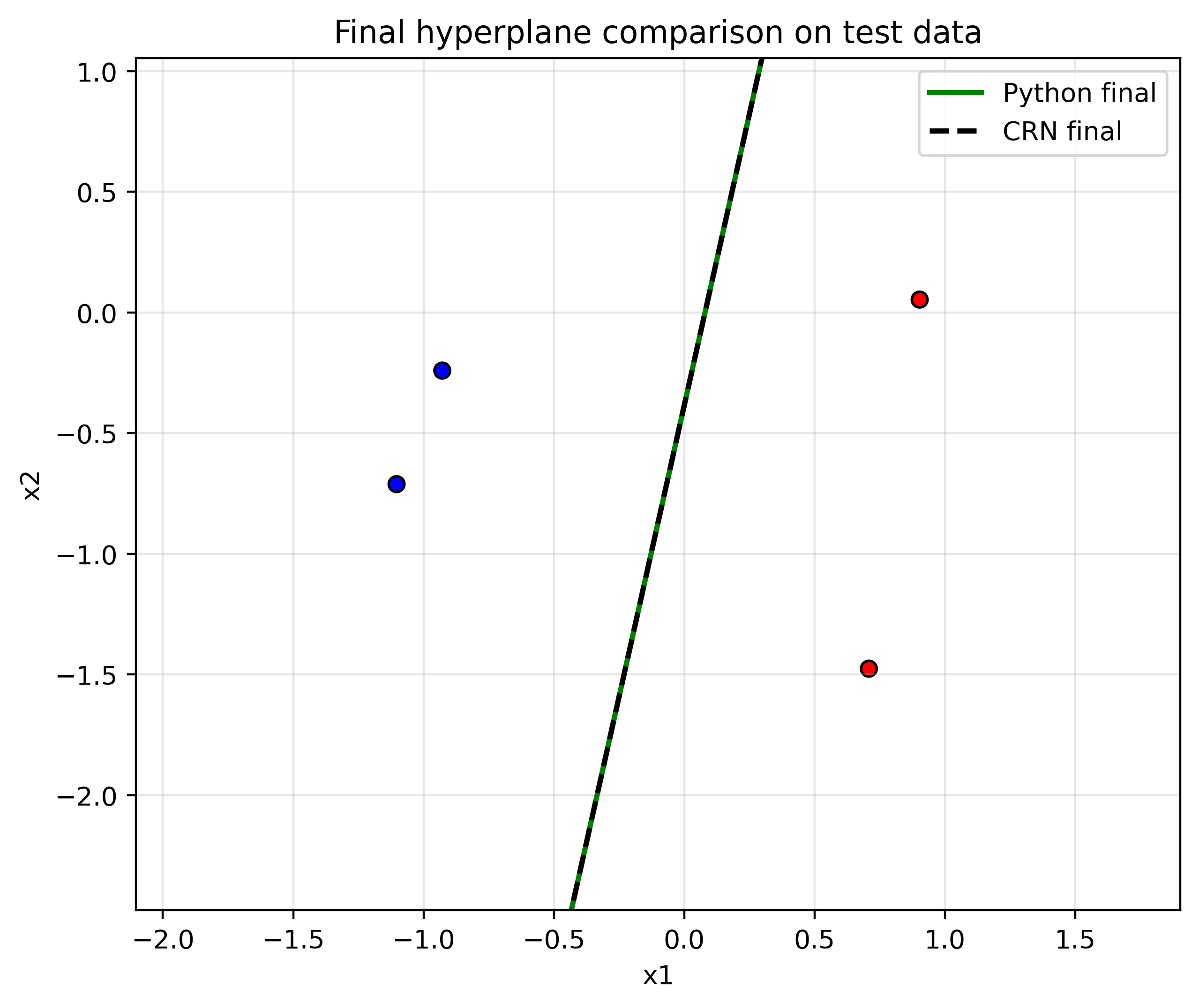}
        \caption{Hyperplane on Test Data}
        \label{fig:test_hyperplane}
    \end{subfigure}

    \caption{Representation of Parameter Evolution and Hyperplane}
    \label{fig:combined}
\end{figure}

To further validate these results, we visualize the evolution of the learned weights and bias throughout the training process. Figure~\ref{fig:combined} provides a consolidated view of the dynamics of the weight parameters, the bias term, and the comparison between the predicted and actual SVM hyperplanes. The plots illustrate how the model's predicted values align with the actual values over the course of {100} training epochs.

\begin{itemize}
{\item Figure \ref{fig:w1} and Figure \ref{fig:w2} (Weight 1 and Weight 2): Initially, the weights exhibit a sharp {incline and} decline respectively, followed by stabilization after approximately {20} epochs.This trend reflects the gradient-based optimization process, where rapid updates occur in the early stages, and convergence is reached as the learning progresses. The predicted values (orange dashed lines) closely track the actual values (solid blue lines), indicating strong predictive accuracy.}

\item Figure \ref{fig:bias} (Bias): The bias shows a similar pattern of rapid early changes before stabilizing.
While slight deviations between the predicted and actual bias exist, the overall trend suggests effective approximation of the learning dynamics.

\item Figure \ref{fig:train_hyperplane} and Figure \ref{fig:test_hyperplane} (Hyperplane Comparison): The actual hyperplane from SVM (green solid line) is compared with the predicted hyperplane (black dashed line).
{In train as well as test case, the hyperplane is able to divide the samples into their respective classes. The CRN hyperplane also  closely follows the traditional hyperplane. The overall similarity suggests that the model effectively captures the underlying decision boundary.}
\end{itemize}

The close alignment between predicted and actual values demonstrates the reliability of the CRN-based predictive model, while also suggesting potential areas for further refinement to enhance precision.

\section{Discussion}\label{sec:discussion}

We have presented a reaction network–based framework for implementing soft-margin support vector machines (SVMs). The proposed method realizes both the feedforward and backpropagation modules of the SVM through a collection of submodules. Specifically, the design incorporates addition, subtraction, multiplication, comparison, and approximate majority modules. Sequential execution of reactions is achieved via time-scale separation, and implemented using an Hopf oscillator. To handle the possibility of negative weights and biases, we utilize dual-rail encoding.

Our work intersects with several other works that have used reaction networks to implement machine learning algorithms. Related approaches include the modeling of neural networks through chemical reaction networks \cite{anderson2021reaction}, the implementation of Boltzmann machines \cite{poole2017chemical,poole2025autonomous}, and the computation of maximum likelihood estimates \cite{gopalkrishnan2016scheme}.

The present study opens multiple avenues for further investigation. One direction is to rigorously analyze the computational complexity of our implementation, in terms of the number of molecular species, the number of reactions, and the convergence rates of the individual computation steps, as in \cite{anderson2025chemical}.
Another direction involves the computational optimality of our design in the sense of \cite{anderson2021reaction}. More specifically, if the reaction network is represented by the dynamical system 
\[
\dot x(t) = f(y,x(t))
\ \text{ with } \
x(t) \in \R^n_{\ge 0}, 
\]
two key questions arise:
\begin{enumerate}
\item 
\textit{Exponential reliability: }
Does there exist a function $\zeta : \mathbb{R}^p_{>0} \to \mathbb{R}_{>0}$ such that
\begin{equation} \notag
|x(t) - \xi(y)| \leq |x(0) - \xi(y)|e^{-\zeta(y) t},
\end{equation}
where $\displaystyle\lim_{t\to\infty}x(t)=\xi(y)$?

\item 
\textit{Finite-time convergence to compact set: }
Do the trajectories converge to a compact set containing the equilibrium in finite time? That is, does there exist a compact set $K \subset \mathbb{R}^n_{\ge 0}$ and a function $T : \mathbb{R}^p_{>0} \to \mathbb{R}_{>0}$ such that $x(t) \in K$ for all $t \geq T(y)$ and $x(0) \in \R^n_{\geq 0}$?
\end{enumerate}
Finally, it is natural to explore extensions of the proposed framework. Possible directions include adapting the scheme to incorporate the \emph{kernel trick} to enable the classification of nonlinear data.

\section{Data accessibility statement}

All relevant codes can be found here: \href{https://github.com/AmeyChoudhary/svm-implementation-using-crn/tree/main}{https://github.com/AmeyChoudhary/svm-implementation-using-crn/tree/main}  

\bibliographystyle{amsplain}
\bibliography{Bibliography}

\newpage

\appendix

\end{document}


\title{Supplementary Information}

\date{}

\author[1]{Amey Choudhary}
\author[2]{Abhishek Deshpande}
\author[3]{Jiaxin Jin}
\affil[1,3]{\small Center for Computational Natural Sciences and Bioinformatics, International Institute of Information Technology, Hyderabad} 
\affil[2]{\small Department of Mathematics, University of Louisiana}

\maketitle

\section{Experimental Setup}\label{sec:experimental_setup}

\subsection{Dataset and Encoding}
We generate a synthetic dataset where each sample consists of:
\begin{itemize}
    \item Input ($X$): A two-dimensional vector.
    \item Output ($Y$): A binary label, either $+1$ or $-1$, indicating class membership.
\end{itemize}

The dataset consists of ten data points distributed across the two classes. To represent numerical values in our computational framework, we use a dual-rail encoding scheme. This encoding transforms each number into two separate values:
\begin{itemize}
    \item A positive component ($x_p$) that holds the original value if positive, or zero otherwise.
    \item A negative component ($x_n$) that holds the approximate value of the number if it is negative, or zero otherwise.
\end{itemize}

The same encoding is applied to the output labels ($Y$) and model parameters.

\subsection{Training Parameters}

We train our CRN-based SVM using the following hyperparameters:
\begin{itemize}
    \item Regularization parameter ($\lambda$): 0.0001
    \item Number of epochs: 500
    \item Learning rate ($\eta$): 0.002
\end{itemize}

\subsection{Computational Framework}

The CRN-based SVM is implemented using a system of ordinary differential equations (ODEs) that simulate biochemical reactions corresponding to arithmetic operations required for SVM optimization. The core operations include:
\begin{itemize}
    \item Addition and subtraction
    \item Multiplication
    \item Division
    \item Comparison and thresholding
\end{itemize}

These operations are implemented as ODEs solved numerically using the \texttt{odeint} function from SciPy. The iterative updates for weight parameters ($w_1$, $w_2$) and bias ($b$) are performed based on the hinge loss function, which is commonly used in SVM optimization.